\documentclass[rsi,aip,twocolumn,amsmath,amssymb]{revtex4}
\usepackage{graphicx}
\usepackage{color}

\begin{document}

\title{A spectroscopic cell for fast pressure jumps across the glass transition line}
\author{R.~Di Leonardo}
\email{roberto.dileonardo@phys.uniroma1.it}
\author{T.~Scopigno}
\author{G.~Ruocco}
\affiliation{Universit\'a di Roma {\rm La Sapienza} and INFM,
I-00185, Roma, Italy.}
\author{U.~Buontempo}
\affiliation{Universit\'a di L'Aquila and INFM, I-67010, L'Aquila, Italy.}
\date{\today}

\begin{abstract}
We present a new experimental protocol for the spectroscopic
study of the dynamics of  glasses in the aging regime induced by
sudden pressure jumps (crunches) across the glass transition
line. The sample, initially in the liquid state, is suddenly
brought in the glassy state, and therefore out of equilibrium, in
a four-window optical crunch cell which is able to perform
pressure jumps of 3 kbar in a time interval of $\approx$10 ms.
The main advantages of this setup with respect to previous
pressure-jump systems is that the pressure jump is induced
through a pressure transmitting fluid mechanically coupled to the
sample stage through a deformable membrane, thus avoiding any
flow of the sample itself in the pressure network and allowing to
deal with highly viscous materials. The dynamics of the sample
during the aging regime is investigated by Brillouin Light
Scattering (BLS). For this purpose the crunch cell is used in
conjunction with a high resolution double monochromator equipped
with a CCD detector. This system is able to record a full
spectrum of a typical glass forming material in a single 1 s
shot. As an example we present the study of the evolution toward
equilibrium of the infinite frequency longitudinal elastic
modulus ($M_\infty$) of low molecular weight polymer
(Poly(bisphenol A-co-epichlorohydrin), glycidyl end capped). The
observed time evolution of $M_\infty$, well represented by a
single stretched exponential, is interpreted within the framework
of the Tool-Narayanaswamy theory.

\end{abstract}

\maketitle

\section{introduction}
Glass forming materials can relax their structure on a timescale which is
dramatically sensitive to control parameters (temperature, pressure, etc.) in
the vicinity of the glass transition region.  When approaching the glass phase
boundary, the typical structural relaxation time increases by several orders of
magnitudes until the system falls out of equilibrium on a macroscopic
timescale.  The physical properties of the so formed glasses depend on history
and evolve with the time spent in the glassy phase. Microscopic dynamics
progressively slows down and the system is said to display physical aging.  The
possibility of recovering a thermodynamic description of glasses in terms of
additional, slowly time dependent, state parameters is still a lively debated
issue in condensed matter physics \cite{davies, McKenna, cugliandolo97, nieu, SciortinoTh, dileo}.  
In this respect, the development of any new experimental protocol capable to probe the 
aging dynamics of glass-forming materials, is crucial to verify and stimulate 
existing theories of the glassy state.
%
%The dynamical properties of glass forming liquids evolving towards
%an equilibrium glassy state (the so called {\it aging} regime) are
%one of the major conundrum in condensed matter physics. The aging
%regime provides a valuable source of information on the physical
%properties of condensed matter. Indeed, during aging, the set of
%thermodynamic variables needed to describe the "state" of the
%system is certainly larger than that usually employed at
%equilibrium (temperature, pressure, etc.) \cite{McKenna}. The
%existence of such an extended parameters space allows to study
%the systems in a much wider number of state points, and, as a
%consequence, the aging regime is thought to be one of the main
%benchmark to test the different existing thermodynamic theories
%of the glassy state \cite{McKenna, SciortinoTh}.
%
%For these reasons the study of
%out-of-equilibrium states has been an active research field since
%the beginning of the last century. 
%As the "aging" regime lasts
%for time comparable to - but not much longer than - the typical
%relaxation times existing in the system, these studies have been
%mainly performed in the glassy phase. In this case indeed the
%relaxation times are very long, ranging from minutes to times
%longer than the experimentalists lifetime, thus allowing to
%easily run the experiments. The main limiting factor for the
%experimental protocols, i.e. the reason that prevents to perform
%"fast" measurements, is the time needed to bring the systems
%off-equilibrium. 
Out of equilibrium states are
customarily produced by fast cooling ({\it quench}). In this case
the cooling time depends on both the quenching rate and the time
required for the sample temperature to equilibrate at the
thermostat temperature.  This latter contribution can be
estimated considering that the thermal diffusivity of a non
metallic liquid is of order $D_T\sim10^{-7} m^2/s$ and that the
typical linear sample dimension in a light scattering experiment
(the one presented here) is $L\sim 1$ cm:
$$\tau_{eq}\simeq\frac{1}{D_T q^2}\simeq 1\;\textrm{min}$$
Even in the (non physical) case of infinite quenching rate, the
above observation implies that one is practically obliged to
quench the sample to temperatures below the glass transition
temperature $T_g$ (where the structural relaxation time
$\tau_\alpha$ reaches $100$ s) and to perform experiments lasting
for hours or even days \cite{macphail}.

The rapid production of non-equilibrium states in glass forming
systems opens, therefore, new possibilities for the spectroscopic
investigation of the aging regime: {\it i)} there is no need to
perform long experiments, the dynamics can be sampled on the
time-scale of seconds, {\it ii)} the aging dynamics for
sufficiently long times does not depend on the detailed
thermodynamic  history.
%(o shape of the transition to the non-equilibrium state).
The investigation of the aging regime over reasonably short
time-scales can be in principle achieved through an alternative
choice of the control parameter, namely performing rapid pressure
jumps ({\it crunches}). In fact, pressure propagates inside the
sample with the speed of sound,  that means:
$$\tau_{eq}\simeq\frac{L}{c}\simeq 10\;\mu\textrm{s}$$
In this conditions what really fixes the crunch time is the
minimum time required to stabilize the pressure on the walls of
the sample cell to the new value, i.e. the experimental crunching
rate.

Pressure jumps have been often used in the past, particularly to
perform studies on chemical reactions kinetic and on protein
unfolding. To our knowledge no attempt has been made to use the
crunches to attain out-of-equilibrium states in disordered
matter. The pressure "jump" (or pressure perturbation as usually
called in the chemical kinetic studies) are used to control the
reactions equilibrium constant via the modulation of the system
free energy, ruled in turn by the pressure. The response of the
system is then detected through the time dependence of the systems
conductivity or optical properties. This method, tracing back to
the works of Maxfield et al.\cite{CleggMaxfield}, is capable of
fast (20 $\mu$s) but moderate (few bar) pressure variation (for a
recent review see \cite{WuLin}). In this type of experiments the
cell is usually a two-windows optical cell, with sapphire windows
and the pressure modulation is attained through a stack of
piezoelectric crystals. When larger pressure variation is needed,
the procedure consists in slowly increasing the pressure of the
liquid sample up to the burst of a safety membrane, which causes
the sudden pressure release. Up to  200 bar in $\approx$100
$\mu$s has been reached with this method \cite{Astumian}.

An "optical" two-windows cell, suitable for light- and
neutron-scattering experiments has been recently employed to
study phase transitions. In this case \cite{Migler} large pressure
"jump" are needed. This is accomplished by controlling the
pressure of a transmitting fluid (silicon oil) separated from the
sample by a deforming gasket (viton o-ring). In this case,
however, due to the mechanical pressure control, the pressure
variation were limited to 30 bar/s. Therefore only "slow"
pressure changes were allowed by this setup.

More recently, an high pressure jump apparatus devoted to X-ray
studies of protein folding has been recently proposed by
Woenckhaus et al. \cite{woe}. This setup is capable of
liquid-liquid pressure jumps with a maximum amplitude of $7$ kbar
on a time scale of $5$ ms. Such a jump is achieved by loading the
sample at high pressure in a storing chamber separated from the
measurement chamber (two-windows sample cell) by a high pressure
valve, then quickly opening the valve. While this scheme can be
profiting exploited in the study of liquid-to-liquid pressure
jump, it presents conceptual shortcomings if one aims to perform
jumps ending up in the glass phase or in a supercooled liquid
phase. Indeed, in the latter cases, the sample in the storage
chamber is already highly viscous, consequently the time needed
to flow throughout valves and pipes is long enough to slow-down
the pressure change.  To overcome this problem a pressure
transmitting fluid -physically separated from the sample- is
mandatory.

The drawback of having the sample in the whole high pressure
network has been overcome by a novel design set-up \cite{StKr}.
This set-up mixes the fast negative pressure jump obtained by
Woenckhaus et al. \cite{woe} (i.e. it makes use a fast high
pressure valve) with the idea of having a transmitting fluid
separated by the sample (in this case by a teflon piston). The
sample cell is a two-windows X-ray cell suitable for small-angle
x-ray scattering, and extensively used to follow the time
evolution of biological samples. The maximum pressure jump is
$-3.5$ kbar and the typical time needed for such jump is $5$ ms,
i.e. the time needed to open the high pressure valve.

In this paper we describe the layout of a new set-up that is
designed to match the following requests: i) four-windows optical
cell to perform light-{\it scattering} experiment (two-windows
cell are not suitable as the back-scattering geometry must be
avoided due to the intrinsic high level of stray light not
compatible with a CCD based detection system, see below); ii)
possibility to perform liquid-to-glass and glass-to-glass
transition, i.e. the sample need to be separated by the pressure
transmitting fluid; iii) initial and final pressure controlled in
the widest possible range ($P_i,\; P_f<7$ Kbar, $\Delta P<2$ Kbar);
iv) quick {\it positive} pressure change ($\approx 5$ ms); v)
possibility of automatic and quick repetition of the "crunches".
This framework is based on a pressure transmitting fluid that
couples a sealed but deforming sample stage to an expansion gas
chamber. In the following we present the details of the device,
along with an example of its application. In particular we study
aging effects on the sound velocity (determined from the dynamic
structure factor measured by Brillouin Light Scattering) of a
glass forming material, Poly(bisphenol A-co-epichlorohydrin),
glycidyl end capped (PBGD), after a pressure jump from the liquid
to the glassy phase close to the glass transition pressure.

\section{The crunch cell}
%\subsection{The crunch cell}

\begin{figure}
\includegraphics[width=.35\textwidth]{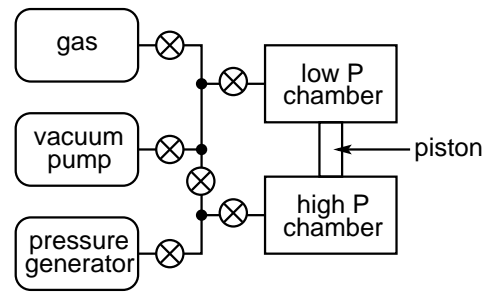}
\caption{ Schematic layout of crunch apparatus. After the
evacuation of the low- and high-P chambers, the gas reservoir is
used to fill the low pressure chamber to the desired pressure and
the pressure generator (a manual Nova Swiss piston) brings the
sample to the initial pressure.}\label{layout}
\end{figure}

The layout of our apparatus consists of two main stages (Fig.
\ref{layout}). A low pressure chamber is loaded with gas
(Nitrogen) through a feed line designed to operate at a maximum
pressure of 700 bar (typical working values are 50 bar).

\begin{figure}
\begin{center}
%\hspace{-2cm}
\includegraphics[width=.5\textwidth]{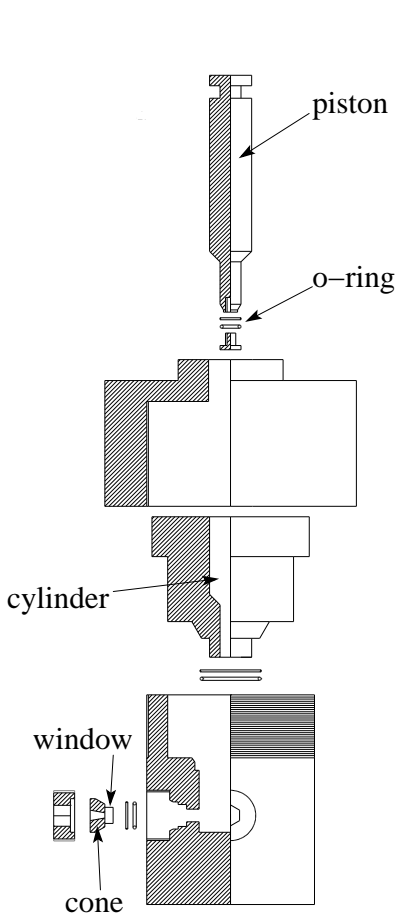}
\caption{ Details of the high pressure chamber. The piston is
mechanically connected to a piston in the low-P chamber (are ratio
1:50) and has a maximum excursion of 10 mm.}  \label{hpc}
\end{center}
\end{figure}

A piston mechanically couples the low pressure stage to a chamber
filled with ethanol (the pressure transmitting medium). This high
pressure stage (Fig. \ref{hpc}) mainly consists of a tempered
steel cell. Four quartz optical windows in right angle geometry
allows for either 90 degrees, forward or back-scattering
geometries. One feed-through allows the pressure transmitting
fluid loading through an high pressure line connected to a
pressure generator Nova Swiss (7 kbar max). Two further feeds
connect an high pressure gauge and a thermocouple. All the
connectors, valves and tubes where purchased from Nova Swiss,
Switzerland.

The two surfaces of the piston are such that the equilibrium position is
reached when the pressure of the fluid in the sample cell is $\sim$ 50 times
bigger than that of the gas. The crunch event is obtained by the following
steps:
\begin{enumerate}
\item an electro-magnet is switched on blocking the piston in its upper position
(smallest volume in the gas chamber, biggest in the transmitting fluid
chamber),
\item the pressure of the transmitting fluid is brought to the desired initial
value ($1\div700\textrm{bar}$) by means of an hand-screw-type  pressure
generator,
\item the gas chamber is filled up to $\sim 60\textrm{bar}$ according to the
final value desired for the pressure in the sample cell (max
$3000\textrm{bar}$),
\item the electromagnet is switched off releasing  the piston which compresses
the transmitting fluid to the final pressure.
\end{enumerate}

\begin{figure}
\begin{center}
\includegraphics[width=.4\textwidth]{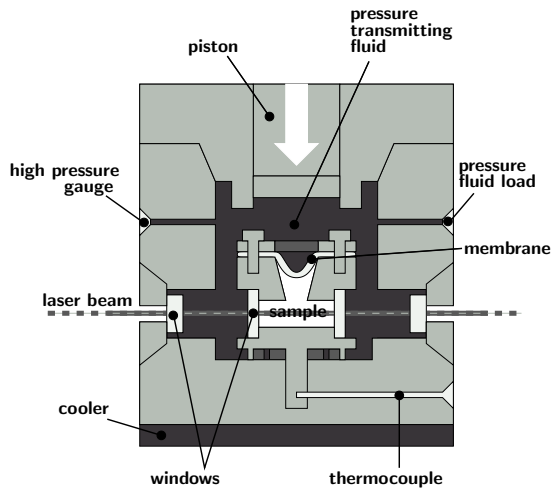}
\caption{ Expansion of the lower part of the four optical windows
high pressure chamber with the sample container (four optical
windows and an elastic membrane to equalize the pressure on the
pressure transmitting fluid and on the sample.} \label{smallcell}
\end{center}
\end{figure}

A small brass cell (Fig. \ref{smallcell}) with a flexible membrane and four optical
quartz windows is filled with sample and immersed in the pressure
transmitting fluid. The membrane mechanically couples the sample
to the pressure transmitting fluid. The pressure of the sample is
monitored indirectly by means of an high pressure strain gauge in
contact with the fluid. The temperature is read on a thermocouple
fixed on the stem of the sample cell.

\begin{figure}
\label{isopentano}
\includegraphics[width=.4\textwidth, angle=-90]{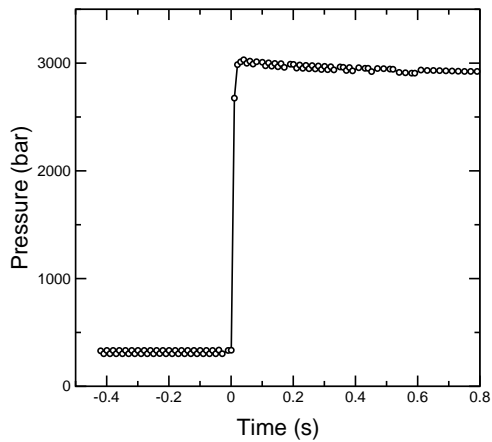}
\caption{Time dependence of the pressure during a crunch event in
a test experiment on iso-pentane. The pressure has been measured
by a pressure (strain) gauge coupled to the transmitting fluid at
a sampling rate of 100 Hz.} \label{isop}
\end{figure}

In standard operation, all the pressure and the temperature gauges
are monitored by means of Gefran 2300 fast display units connected
to the serial port of a PC. The maximum recording rate for
pressure is $100\textrm{Hz}$. In the commissioning phase, the
pressure gauge is read via a digital oscilloscope in order  to
trace the pressure change in the sample with the desired time
resolution. A typical pressure profile of a test fluid
(iso-pentane) in the sample chamber during a crunch event is
reported in Fig.~\ref{isop}. As can be noticed, in this example
the pressure jumps from $\sim 300 \textrm{bar}$ up to $\sim 3000$
in $\sim$ 10 ms and the  temperature changes by $\approx$2 K
(strongly sample dependent). The subsequent slow ($\sim$ 0.5 s)
decrease of pressure is due to the thermalization of the fluids
to room temperature. The low compressibility of the gas (as
compared to those of the pressure transmitting fluid and of the
sample) guarantees the pressure stability against small volume
rearrangement due to the (expected) slow time dependence of the
compressibility in the aging sample. The latter point is
important as, with respect to other set-ups , the present device
is designed to work at {\it constant} pressure, even in presence
of relaxation in the physical properties of the sample.

\section{An application: aging of the dynamic structure factor}

\subsection{Experimental details and results}

As an example of application of the crunch apparatus we present
the investigation of aging effects in the dynamics of
Poly(bisphenol A-co-epichlorohydrin), glycidyl end capped
(PBGD).  In particular we monitor the time evolution of the
infinite frequency longitudinal elastic modulus $M_{\infty}$
(defined as the instantaneous part of pressure response to an
infinitesimal relative volume variation) as determined from the
dynamic structure factor obtained by Brillouin Light Scattering
(BLS). We will discuss the results in the framework of the old
fictive pressure concept firstly introduced by Tool and
co-workers \cite{TN}. The Brillouin light scattering spectra are
measured by a CCD- and monochromator-based system. The incoming
beam, $500$ mW at $\lambda$=514.5 nm ,  from  a Coherent INNOVA
Ar-ion laser operating in single mode owing to an intra-cavity
ethalon and in power-stabilized mode, was focused on the sample.
The polarization of the incident beam is vertical with respect to
the scattering plane. The light, scattered at an average angle of
$ 90ø $, was collected by a field lens ($20$ cm focal length, 4
cm diameter).
%the finite dimensions of which give an angular acceptance of XXX
%sr, leading to a leakage of polarized scattered light in the
%depolarized spectra of the order of XXX of the total intensity.
After selecting  the vertical polarization by a polaroid film
(rejection $10^{-4}$), the  scattered light is focused on the
entrance slit (20 $\mu$m in the dispersion plane $\times$ 2 mm)
of  a SOPRA DMDP2000 monochromator. The latter is composed of a
couple of two meter focal length grating monochromators, in
Fastie-Ebert mounting, each one with entrance and exit slits and
coupled in additive dispersion by an external 1:1 telescope. The
monochromator was operating in the single pass/double
monochromator (DMSP) configuration \cite{sopra} at the 11-th
diffraction order, corresponding to a transmissions of $\approx
25 \%$ and a total dispersing power on the exit slit plane equal
to 1 cm$^{-1}$/mm. In a conventional experiment all the slits are
closed according to the desired resolution width, a
photo-multiplier is placed in front of the exit slit and
frequency scans of scattered light are obtained by simultaneous
rotation of the two gratings. In this operating mode a full
Brillouin spectrum with good statistic is acquired in $30-60$
minutes. For the reasons discussed in the introduction one aims
to reduce the observation time as much as possible, to tens of ms
having in mind that the crunch event lasts for $\approx 10$ ms.
In this respect the use of a CCD detector turns out to be an
appropriate choice as it allows the detection of a full Brillouin
spectrum, with a good statistics, in less than one second.
Therefore, in our setup, the two intermediate slits and the exit
slit of the monochromator are fully open (2 mm $\times$ 2 mm) and
the image formed in the final plane slit is focused via of a
single achromat doublet (focal length 75 mm, magnification
$\sim10$) on the surface of a CCD camera. The choice of the
focusing lens has been done in order to maintain aberrations well
below the monochromator's resolution. The camera is a HiResIII
camera (DTA, Pisa, Italy) mounting a SITe 1100x330 pixel (total
dimension 2 mm $\times$ 0.6 mm) back-illuminated CCD sensor
(quantum efficiency of $\sim 80\%$) cooled via a Peltier element
and a fluid close circuit system based on an HAAKE 75
refrigerator. In this set up, the spectral range covered by the
CCD sensor is 2 cm$^{-1}$, i.e. 0.002 cm$^{-1}$/pixel, that
correspond to ten points on each spectral resolution as dictated
by the width of the monochromator entrance slit. The image of the
camera is transferred to a PC, where the BLS spectrum is obtained
by integrating the image in vertical (non-dispersing) dimension,
and by performing the usual operation (background subtraction,
flat field correction, etc.)

\begin{figure}
%\centering
%\vspace{-.2cm}
%\hspace{-.6cm}
\includegraphics[width=.45\textwidth,angle=0]{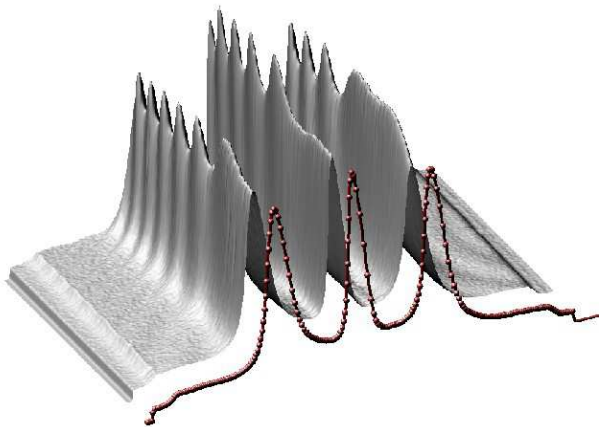}
\caption{Three-dimensional image of the intensity measured by the
CCD detector (exposure time $10 s$) on a sample of toluene. The
surface represents the light intensity while the line is the
average intensity over CCD rows orthogonal to the dispersion
direction of the spectrometer. The intensity oscillation are due
to interference in the CCD silicon wafer. The intensity drops in
the far tails of the Brillouin triplet is due to the spectral cut
of the 1 mm width intermediate slits of the spectrometer.}
\label{toluene}
\end{figure}

As an example, in Fig.~\ref{toluene} we show the Brillouin
spectrum of Toluene in a $90^\circ$ scattering geometry at
ambient temperature and pressure. The exposure time is only 10 s,
but the statistical quality of data is comparable to that of a
1000 s long frequency scan performed with photo-multiplier. In
the 3-D plot a fringes pattern due to interference of coherent
light reflected by the (non-wedged) CCD array surface is clearly
visible. The presence of such a fringe pattern is not relevant in
the present application as the final information is obtained
integrating the image in the vertical dimension (full line in the
figure) and, therefore, the intensity oscillation introduced by
the interference is averaged out. The spectra is made by the
well-know triplet of lines. The central one brings contribution
from the entropic mode (thermal diffusion process) and from the
structural relaxation (Mountain peak, non visible here as the
toluene at room condition has a relaxation time much shorter than
the inverse of the frequencies covered in Fig.~\ref{toluene}. The
side peaks are due to the pressure modes, and, in particular,
their frequency position and width bring information on the
elastic moduli and on the liquid viscosity. Equilibrium spectra
like that in Fig.~\ref{toluene} are used to calibrate the
frequency scale on the CCD surface, i.e. the pixel-to-wavenumber
conversion factor.

A second calibration, which will turn out useful in the data
analysis, is the relation existing, for the specific investigated
sample, between the applied pressure and the elastic modulus for
the liquid {\it at equilibrium}.

\begin{figure}[t]
\includegraphics[width=.45\textwidth]{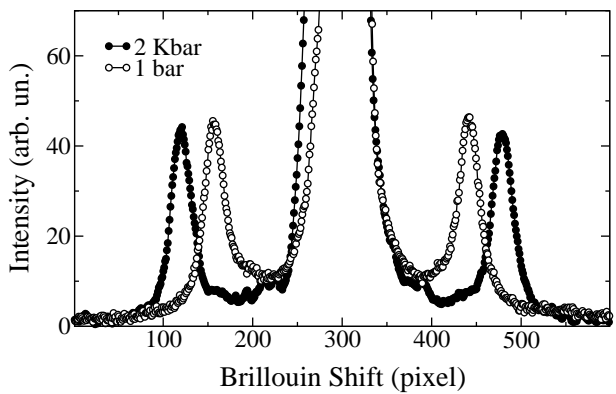}
\caption{Example of Brillouin spectra of (equilibrated) PBGD at
room pressure (open circles) and at 2 Kbar (full circles) in the
crunch cell. The intense elastic line is due to both parasitic
light coming from the optical windows and from the mountain mode
(which become quasi-elastic in the investigated thermodynamic
conditions) of PBGD.}\label{PBGD_Eq}
\end{figure}

In Fig.~\ref{PBGD_Eq} we report as an example the spectrum of PBGD
at room temperature and $P$=1, 2000 bar. As in the case of
toluene, the spectra shows the three peaks structure of Brillouin
spectra of liquids. In this case the central component is the sum
of three unresolved contributes i) the Rayleigh line: light
scattered by the slow decaying component of density fluctuation
driven by thermal diffusion ($\tau\sim 1\mu s$); ii) the Mountain
line: light scattered by the slow decaying component of density
fluctuation driven by structural relaxation ($\tau> 1\mu s$);
iii) the Stray light: light scattered elastically  by particles
in the sample and optical components (windows, lenses, etc.). In
addition to the central component the spectra display also a
clearly visible Brillouin  doublet arising from acoustic damped
oscillations in the density fluctuation decay. Since the
frequency of these oscillations is of order $10$ GHz all the
previously mentioned slow modes are basically frozen on the
acoustic timescale the peak position ($\omega_\infty$) is
obtained from the infinite frequency modulus:
$$\omega_{\infty}^2=\frac{q^2 n^2}{\rho} M_{\infty}.$$
$$M_{\infty}=V \left (\frac{\partial P}{\partial V}\right )_{S,\xi}$$
Where the subscripts $S$ and  $\xi$ mean that the thermodynamic
derivative has to be computed for constant entropy (frozen
thermal diffusion) and fixed structure (frozen structural
relaxation). Since the variation of $n^2/\rho$ with pressure turns
out to be much more weak than that of $M_{\infty}$,
$\omega_{\infty}^2$ is basically proportional to $M_{\infty}$.

We have fitted the Brillouin peaks with two Lorentzians in order
to get information on the peak position and indirectly on
$M_{\infty}$. These equilibrium spectra have been measured from
$P$=0 to $P$=2.9 kbar, each time waiting for the system to
equilibrate before performing the data collection. It is worth
noticing that in this pressure range the relaxation time of PBGD
(as determined by dielectric/PCS spectroscopy, \cite{paluchprl})
are in the 0.1$\div$100 ms range.

\begin{figure}[t]
\includegraphics[width=.3\textwidth]{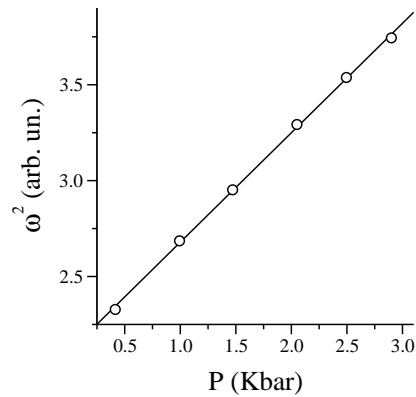}
\caption{\label{omeq} Squared frequency of Brillouin peaks
($\circ$) as function of pressure for equilibrated PBGD. The
solid line represents the best fit with a linear law}
\end{figure}

%What we can actually access in this kind of experiment is the
%infinite frequency Brillouin shift $\omega_{\infty}$ which is
%related to the unrelaxed longitudinal elastic modulus $M_{\infty}$ by:
The calibration measurement, reported in Fig.~\ref{omeq},
indicates that $\omega_{\infty}^2(P)$ is linearly proportional to
the pressure in the investigated pressure range. This result, as
we will see later, simplifies the data analysis however is not
crucial. Indeed, even in the case of non-linear relation between
moduli and pressure, the calibration curve can be used to extract
the time evolution of the (apparent) pressure from that of the
moduli.

\begin{figure}
\hspace{-.3cm}
\includegraphics[width=.45\textwidth]{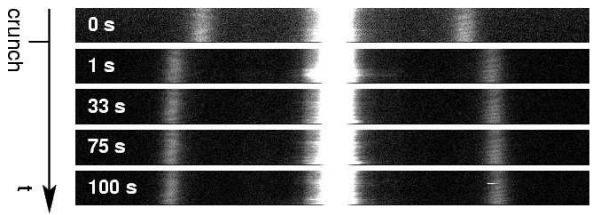}
\caption{CCD images of Brillouin spectra of PBGD at different
waiting times during the aging dynamics. The intense central peak
and the two weaker side peaks are clearly visible. Time zero
indicate the spectra taken just before the crunch. Soon after the
crunch the Brillouin peaks jumps to a new value, which then
slowly changes during the aging process.} \label{ccdcru}
\end{figure}

\begin{figure}
\hspace{1cm}
\includegraphics[width=.4\textwidth]{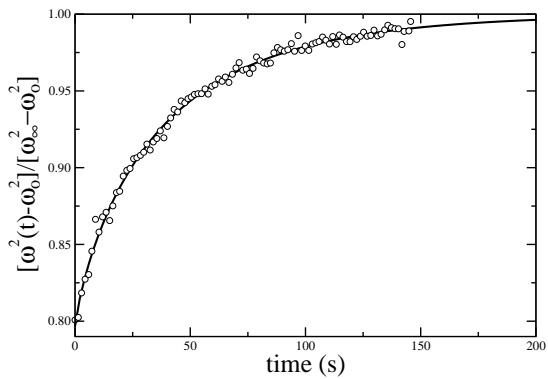}
\caption{Plot of the infinite frequency modulus (open circles) as
a function of waiting time $t_w$ after the crunch. The initial
80\% of the change in $\omega^2$ is due to the "vibrational"
component rearrangement, the residual 20\% -shown here- is the
slow structural rearrangement. The solid line is a fit to the
expected stretched exponential behavior with $\tau,\beta$ free
parameters. }\label{fitcru}
\end{figure}

As an example of crunch experiment, in Fig. \ref{ccdcru} we report
a (selection of) CCD images with exposure time $1$ s following a
pressure jump from $0.6$ to $2.6$ kbar. Each image displays the
three peaks structure of Brillouin spectra of liquids. We have
fitted the Brillouin peaks with two Lorentzian lines in order to
get information on the peak position and indirectly on
$M_{\infty}$. The obtained values, reported as a function of
time in Fig. \ref{fitcru}, are very well represented by a stretched
exponential law. In the following section we will show that such a stretched
exponential behavior is consistent with the prediction of
the Tool-Narayanaswamy model in the approximation of a very fast crunch.

\begin{figure}
  \begin{center}
    \includegraphics[width=.3\textwidth]{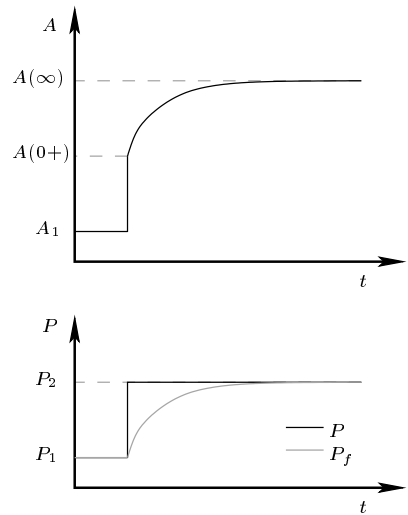}
  \end{center}
\caption{The upper panel reports a sketch of the expected time
evolution of a generic macroscopic observable $A$ after a crunch
from $P_1$ to $P_2$. The time dependence of the external (thick
line) and fictive (thin line) pressures are qualitatively reported
in the lower panel.}
  \vspace{.5cm}
  \label{A_t}
\end{figure}

\subsection{The Tool-Narayanaswamy model}
Consider a supercooled liquid, initially at equilibrium, whose
pressure is abruptly changed from $P_1$ to $P_2$. The time
evolution of a generic macroscopic observable $A$ is sketched in
Fig.~\ref{A_t}. As rapidly as the pressure is changed $A$ jumps
from $A_1$ (equilibrium value at $P=P_1$) to $A(0+)$ in a
glass-like fashion, i.~e. similarly to what would have been
happened if the $P_1 \rightarrow P_2$ jump was taking place in
the deep glassy state, when only the vibrational degrees of
freedom are free to rearrange and the structure is frozen. This
initial, adiabatic, modification of $A$ is followed by a slow
structural relaxation, during which the observable $A$ changes
with time from $A(0+)$ to the new equilibrium value $A(\infty)$.
Let's call $\phi$ the relaxational part of $A(t)$:
$$\phi(t)=\frac{A(\infty)-A(t)}{A(\infty)-A(0+)}$$
The "fictive pressure"  is a variable introduced for
computational and conceptual convenience and represents $\phi$ in
units of pressure:
\begin{equation}
  P_f=P_2-(P_2-P_1) \phi(t) \label{Pf}
\end{equation}
The relevance of the concept of "fictive pressure" lies in the
observation that, while we know that at equilibrium $A$ is a well
defined function of pressure and temperature, we argue that out
of equilibrium the state of the system is specified only if we
know the additional state variable $P_f$. This quantity evolves
from $P_1$ to $P_2$, reflecting the slow evolution of the
system's structure towards equilibrium.
\begin{figure}
  \begin{center}
    \includegraphics[width=.3\textwidth]{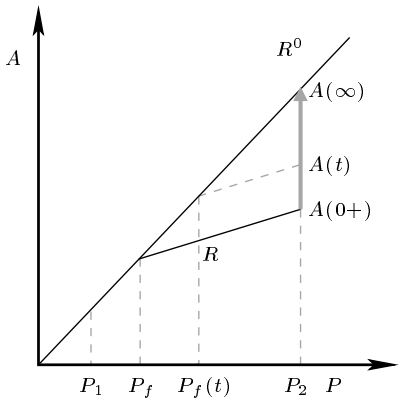}
  \end{center}
\caption{Expected pressure dependence of a generic observable
$A$.}\label{A_P}
\end{figure}

In those cases where $A$ depends linearly on pressure $P_f$
assumes a more physical interpretation. Let's consider an
equilibrated liquid system at initial pressure $P_1$, compressed
with a finite compression rate $R=dP/dt$. Initially, the
equilibrium relaxation time is so short that the system will
adiabatically follow the pressure change, and the variable $A$
evolves along its equilibrium curve (from $P_1$ to $P_f$ in
Fig.~\ref{A_P}). At a given pressure ($P_f$) the relaxation time
becomes so long that the system falls out of equilibrium, and the
variable $A$ follows a weaker (solid-like, here assumed linear
and not depending on $R$) $P$ dependence (from $P_f$ to $P_2$).
The so defined $P_f$ is a function of the compression rate $R$ and
in particular will be greater the slower is $R$. The value
$A(0+)$ reached after the compression will also depend on $R$ or,
equivalently, on $P_f$: $A(0+)=A(T,P_2,P_f)$. Inverting the last
equality we obtain $P_f=P_f(T,P_2,A)$. Once the final pressure is
reached, and $A$ is equal to $A(0+)$, the aging process starts.
$A$ evolves toward $A(\infty)$ and to each $A(t)$ we can
associate a fictive pressure $P_f(t)$ following the dashed line
in Fig.~\ref{A_P} which is the one defined in (\ref{Pf}). If the
so defined fictive pressure is a true state variable
\footnote{$P_f$ is solely defined by $T$,$P_2$ and $A$
independently from the system's history.} the value of $P_f$
obtained from (\ref{Pf}) has the special meaning of that pressure
where the system  would fall out of equilibrium in a compression
experiment leading directly to $A(t)$. 
%In other words, if at time
%$t$ during the aging process we release the pressure from $P_2$
%to $P_f(t)$ we would reach an equilibrium state without further
%aging.

The structural relaxation process in the aging regime is both
non-exponential and non linear. The most widely applied treatment
of this features is embodied in what is called the
Tool-Narayanaswamy (TN) model \cite{TN}, which postulates the
following expression for $\phi(t)$:
\begin{equation}
  \phi(t)=\sum_i g_i \exp\left[-\int_0^t
  \frac{dt'}{\tau_i(P,P_f)}\right] \label{TNeq}
\end{equation}
The non-exponential character of the relaxation is accounted for
by invoking a distribution of relaxation times $\tau_i$ with
weights $g_i$. The non-linear character is accounted for by
allowing the relaxation times to depend both on pressure, $P$,
and on the average structural state of the system as measured,
for example, by $P_f$. The number of adjustable parameters
entering in this model can be considerably reduced if one assumes
a continuous spectrum of relaxation times, which shape is
specified by a single, pressure independent parameter and its
location on a logarithmic timescale is specified by a reference
relaxation time $\tau$ ($\tau_i=\tau/\lambda_i$). The most common
choice for the spectrum of relaxation times is the
Kohlrausch-Williams-Watts (KWW) or stretched exponential function:
\begin{equation} \label{KWW}
  \sum_i g_i \exp\left(-\frac{t}{\tau_i}\right)=
  \sum_i g_i \exp\left(-\lambda_i\frac{t}{\tau}\right)=
  \exp\left[-\left(\frac{t}{\tau}\right)^\beta\right]
\end{equation}
On a general ground, the relaxation in response to a complicated
pressure history, e.g., a pressure jump $\Delta P_1$ at time
$t_1$, $\Delta P_2$ at $t_2$, ...$\Delta P_m$ at $t_m$, may be
dealt with by invoking the superposition principle \cite{TN}, so
that the fictive pressure $P_f(t)$ at time $t>t_m$ is given by:
\begin{equation}
  P_f(t)=P_1+\sum_{j=0}^m \Delta P_j [1-\phi(t-t_j)]
\end{equation}
or in integral form:
\begin{equation}
  P_f(t)=P_1+\int_0^t R(t')\left[1-\phi(t-t')\right]dt' \label{Pf_t}
\end{equation}
where $R=dP/dt$ is the compression rate.
In order to solve (\ref{Pf_t}) we need an explicit form for $\tau(P,P_f)$.
The simplest choice comes from the observation that for short enough
pressure intervals $\tau(P)$ for an isothermal equilibrated sample is
well described by\footnote{See \cite{paluchprl} for a detailed discussion of
  temperature and pressure dependence of structural relaxation in
  fragile glass formers}:
\begin{equation} \label{tau_P}
  \tau_{eq}(P)=\tau_0\exp(\alpha P)
\end{equation}
The simplest generalization is that of replacing $P$ in (\ref{tau_P})
with a linear combination of $P$ and $P_f$ \cite{montrose}:
\begin{equation} \label{tau_P_oe}
  \tau=\tau_0 \exp\left[\alpha(xP+(1-x)P_f)\right]
\end{equation}
which reduces to (\ref{tau_P}) in the equilibrium case ($P_f=P$).

\subsection{Evolution of fictive pressure in the crunch approximation}
\label{crunchapprox}

If we crunch a system from $P_1$ to $P_2$ the subsequent
evolution of fictive pressure will depend on the pressure profile
as expressed by (\ref{Pf_t}). In these circumstances (\ref{Pf_t})
has to be integrated numerically \cite{bezot}. However if the
compression is fast enough $R(t)$ in (\ref{Pf_t}) can be replaced
by $(P_2-P_1)\delta(t)$ and the detailed shape of the pressure
profile during the compression becomes irrelevant:
\begin{equation}
  P_f(t)=P_2-\Delta\phi(t)\label{Pf_t_c},\;\;\; \Delta=P_2-P_1
\end{equation}
Substituting (\ref{TNeq}) and(\ref{KWW}) in (\ref{Pf_t_c}) we obtain:
\begin{equation}
  \phi(t)=\frac{P_2-P_f(t)}{\Delta}=\exp\left[-\left(\int_0^t
      \frac{dt'}{\tau(P_2,P_f(t'))}\right)^\beta\right]
\end{equation}
and by inserting (\ref{tau_P_oe}):
\begin{eqnarray}
  \phi(t)&=&\exp\left[-\left(\frac{\xi(t)}
      {\tau_{eq}(P_2)}\right)^\beta\right]\\\label{ficru1}
  \xi(t)&=&\int_0^t \exp\left[\bar{\alpha} \phi(t')\right] dt'\label{ficru2}
\end{eqnarray}
where the dimensionless parameter  $\bar{\alpha}$ is equal to
$\alpha (1-x) \Delta$. Equation (\ref{ficru1}) cannot be solved
analytically but we provide here an approximated solution which
holds in an easy estimable range of times. We start by assuming
that $\phi(t)$ is well represented by the stretched exponential:
\begin{equation}\label{fiapp}
  \phi(t) \simeq \exp\left[-\left(\frac{t}{\tau^*}\right)^{\beta^*}\right]
\end{equation}
In the logarithmic time interval $|\ln(t)-\ln(\tau^*)|\ll1/\beta^*$
one can replace the stretched exponential in (\ref{fiapp}) with:
\begin{equation}\label{svilup}
  \exp\left[-\left(\frac{t}{\tau^*}\right)^{\beta^*}\right]\simeq\frac{1}{e}
  \left(\frac{t}{\tau^*}\right)^{-\beta^*}
\end{equation}
which substituted in (\ref{ficru2}) and using (\ref{svilup}) once again
gives:
\begin{equation}
  \label{}
  \xi(t)=at^b, \;\;\; b=1-\frac{\bar{\alpha}\beta^*}{e},
  \;\;\; a=\frac{(e{\tau^*}^{\beta^*})^{\frac{\bar{\alpha}}{e}}}{b}
\end{equation}
Substituting the above expression for $\xi(t)$ in (\ref{ficru1})
and equating the result to (\ref{fiapp}) we obtain $\tau^*$ and $\beta^*$
as functions of $\tau_{eq}(P_2)$, $\beta$ and $\bar{\alpha}$:
\begin{eqnarray}
  \label{taubeta}
  \tau^*=\tau_{eq}(P_2)
  \frac{e^{-\frac{\bar{\alpha}}{e}}}{1+\frac{\bar{\alpha}\beta}{e}}\;\;,\;\;\;\;
  \beta^*=\frac{\beta}{{1+\frac{\bar{\alpha}\beta}{e}}}
\end{eqnarray}
This result is based on the approximation (\ref{svilup}) which
holds in a wider time range the smaller is $\beta^*$. Since
$\beta^*$ decreases with increasing $\bar{\alpha}=\alpha (1-x)
\Delta$ we can conclude that for big pressure jumps ($\Delta$) we
expect, on the basis of TN model, that the aging dynamics of a
generic observable, linearly dependent on $P_f$, could be
described by  stretched exponential with parameters
$\tau^*,\beta^*$ in Eq.~(\ref{taubeta}).

A fit (Fig.~\ref{fitcru}) to the experimental data with a
stretched exponential function with $\tau,\beta$ as free
parameters shows indeed a very good qualitative agreement with
the TN model predictions. It is not our aim here to discuss the
physical meaning of these parameters, we want only to stress the
importance of making experiments with a high compression rate
that lead to easy-to-interpret results (see Eq.~\ref{taubetai}).

\section{Conclusions}
We presented an experimental device able to perform crunch
events, specifically fast pressure jumps of the order of 3 Kbar,
in a time interval shorter than 10 ms. The sample is confined in a
small environment which is mechanically coupled to the pressure
transmitting fluid by a flexible membrane. Such a configuration
allows the possibility of performing fast crunches into highly
viscous phases opening the way to investigations of aging
dynamics on the timescale of seconds. An experimental setup,
based on the crunch cell and on a double grating monochromator
equipped with a CCD detector, has been developed and used to
collect Brillouin spectra with a repetition rate of $\sim 1$ s
during an aging regime with characteristic timescale of $\approx
100 s$. The aging of the dynamic structure factor of s typical
polymeric glass former has been studied and, in particular, the
relaxation of the infinite frequency longitudinal modulus has
been reported as a function of waiting time. A stretched
exponential behavior has been found and justified according to
the Tool-Naranayaswamy fictive pressure model. The systematic
study of nonlinear modulus relaxation for pressure jumps of
increasing entity will be performed in order to verify or discard
the prediction obtained in Sec. \ref{crunchapprox}.


\begin{references}

\bibitem{davies}
R.O. Davies, G.O. Jones, Adv. Phys. {\bf 2}, 370 (1953).

\bibitem{McKenna}
G.~B.~McKenna in Comprehensive Polymer Science,
Vol. 2, Polymer Properties, ed. by C.~Booth and C.~Price,
Pergamon, Oxford (1989), p 311.

\bibitem{cugliandolo97}
L.F. Cugliandolo, J. Kurchan, L. Peliti, Phys. Rev. E {\bf 55}, 3898 (1997).

\bibitem{nieu}
Th.M. Nieuwenhuizen, Phys. Rev. Lett. {\bf 80}, 5580 (1998).

\bibitem{SciortinoTh}
S. Mossa, E. La Nave, F. Sciortino, and P. Tartaglia
Eur. Phys. J. B {\bf 30}, 351 (2002).

\bibitem{dileo}
R. Di Leonardo, A Taschin, R. Torre, M. Sampoli, and G. Ruocco,
Physical Review E {\bf 67}, 015102 (2003). 

\bibitem{macphail}
R.~S.~Miller and R.~A.~MacPhail, J. Chem. Phys., {\bf 106}, 3393,
(1997).

\bibitem{CleggMaxfield}
R.~M.~Clegg and B.~W.~Maxfield, Rev. Sci. Instrum., {\bf 47},
1383, (1976).

\bibitem{WuLin}
C.-H.~Wu, C.-F.~Lin, S.-L.~Lo and T.~Yasunaga, Proc. Natl. Sci.
Counc., ROC(A), {\bf 23}, 466, (1999).

\bibitem{Astumian}
R.~D.~Astumian, M.~Sasaki, T.~Yasunaga and Z.~A.~Schelly, J.
Phys. Chem., {\bf 85}, 3832, (1981).

\bibitem{Migler}
K.~B.~Migler and C.~C.~Han, Macromolecules, {\bf 31}, 360, (1998).

\bibitem{woe}
J. Woeckhaus, R. K\"ohling, R. Winter, P. Thiyagarajan, and S.
Finet, Rev. Sci. Instr. {\bf 71} 3895 (2000).

\bibitem{StKr}
M.~Steinhart, M.~Kriechbaum, K.~Pressl, H.~Amenitsch, P.~Laggner,
and S.~Bernstorff, Rev. Sci. Instrum., {\bf 70}, 1540, (1999).

\bibitem{TN}
O.S.~Narayanaswamy, J. Am. Ceram. Soc.{\bf 54}, 491, (1971)

\bibitem{sopra}
P. Benassi, V. Mazzacurati, G. Ruocco, Eng. Optics {\bf 11}, 533
(1988); ibid. J. of Phys. E {\bf 21}, 798 (1988).

\bibitem{paluchprl}
M.~Paluch, A.~Patkowski, and E.W.~Fischer, Phys. Rev. Lett, {\bf
85}, 2140, (2000).

\bibitem{montrose}
D.M.~Heyes and C.J.~Montrose, Trans. ASME F {\bf 105}, 280, (1980)

\bibitem{bezot} P.~Bezot and C.~Hesse-Bezot, J. Non-Cryst. Solids {\bf 122}, 160 (1990).

\end{references}
\end{document}